\documentclass[12pt]{article}
\usepackage{graphicx}
\usepackage{natbib} 
\usepackage{url} 

\newcommand{\blind}{0}

\usepackage{authblk}
\usepackage{amsmath}
\usepackage{amssymb}
\usepackage{tikz}
\usetikzlibrary{arrows}
\usepackage{float} 
\usepackage{subcaption}
\usepackage[toc,page]{appendix}

\usepackage{booktabs}
\usepackage{algorithm}
\usepackage{algpseudocode}
\usepackage{bbm}
\usepackage{adjustbox}
\newtheorem{prop}{Proposition}
\newcommand{\bi}{\begin{itemize}}
\newcommand{\ei}{\end{itemize}}
\newcommand{\bs}[1]{\boldsymbol{#1}}

\usepackage[shortlabels]{enumitem}

\usepackage{caption}
\makeatletter
\g@addto@macro\normalsize{%
  \setlength\abovedisplayskip{3pt}
  \setlength\belowdisplayskip{3pt}
  \setlength\abovedisplayshortskip{3pt}
  \setlength\belowdisplayshortskip{3pt}
}
\makeatother

\begin{document}

\def\spacingset#1{\renewcommand{\baselinestretch}%
{#1}\small\normalsize} \spacingset{1}


\if0\blind
{
  \title{\bf QuIP: Experimental design for expensive simulators with many Qualitative factors via Integer Programming}
  \small
   \author[1]{Yen-Chun Liu}
    \author[1]{Simon Mak}
    \affil[1]{Department of Statistical Science, Duke University}
  \maketitle
} \fi

\if1\blind
{
  \bigskip
  \bigskip
  \bigskip
  \begin{center}
    {\LARGE\bf }
\end{center}

  \medskip
} \fi

\bigskip
\begin{abstract}

The need to explore and/or optimize expensive simulators with many qualitative factors arises in broad scientific and engineering problems. Our motivating application lies in path planning -- the exploration of feasible paths for navigation, which plays an important role in robotics, surgical planning and assembly planning. Here, the feasibility of a path is evaluated via expensive virtual experiments, and its parameter space is typically discrete and high-dimensional. A carefully selected experimental design is thus essential for timely decision-making. We propose here a novel framework, called QuIP, for experimental design of Qualitative factors via Integer Programming under a Gaussian process surrogate model with an exchangeable covariance function. For initial design, we show that its asymptotic D-optimal design can be formulated as a variant of the well-known assignment problem in operations research, which can be efficiently solved to global optimality using state-of-the-art integer programming solvers. For sequential design (specifically, for active learning or black-box optimization), we show that its design criterion can similarly be formulated as an assignment problem, thus enabling efficient and reliable optimization with existing solvers. We then demonstrate the effectiveness of QuIP over existing methods in a suite of path planning experiments and an application to rover trajectory optimization.


\end{abstract}

\noindent%
\vfill

\newpage
\spacingset{1.7} 
\section{Introduction}

With breakthroughs in scientific computing methods, algorithms and architecture, virtual simulations are quickly replacing physical experiments in broad scientific and engineering disciplines. Highly complex phenomena, e.g., universe expansions \citep{kaufman2011efficient}, particle collisions \citep{ji2023graphical,ji2022multi} and human organs \citep{chen2021function}, can now be reliably simulated via virtual experiments, leading to significant scientific advancements. One bottleneck, however, is that these simulations can be computationally costly, requiring thousands of CPU hours per run. Such costs can thus hamper the exploration and/or optimization of these simulators over a desired parameter space. This is further compounded by the presence of many qualitative factors of interest, which may represent, e.g., policy choices, design decisions or control actions. Many such parameters can bring about a combinatorial explosion of the parameter space, making its exploration highly challenging. 

In this paper, we target the motivating application of path planning, which aims to find an optimal path through difficult terrain. Path planning plays an important role in the development of, e.g., unmanned aerial vehicles \citep{phung2017enhanced,narayanan2024misfire}, spacecraft \citep{aoude2008two} and robotics \citep{shareef2016simultaneous}. In practice, the decision space for path planning is often discretized to many qualitative variables, each representing a path decision over a time step. The optimization of such decision variables is known as discrete path planning \citep{gasparetto2015path,karur2021survey}. This is particularly useful when the vehicle in motion is restricted to a finite set of actions, e.g., due to engineering constraints. For simple applications where path planning is cheap, variants of Dijkstra's algorithm \citep{dijkstra1976discipline}, rapidly-exploring random trees \citep{lavalle2001rapidly} and particle swarm optimization \citep{kennedy1995particle} have been employed. For complex applications, however, expensive simulations are needed to faithfully represent the complex multi-physics environment \citep{li2023additive}. For example, a realistic simulation of a single spacecraft trajectory can require tens of CPU hours to perform \citep{murman2005characterization, hughes2016general}. The optimization of such routes over a high-dimensional combinatorial space can thus be highly challenging.

One promising solution is surrogate modeling. The idea is to train a probabilistic predictive model that can efficiently emulate the expensive simulator with uncertainty. Gaussian processes (GPs; \citealp{gramacy2020surrogates}) are widely used as surrogates; they provide closed-form expressions for prediction and uncertainty quantification, which enable efficient downstream decision-making. There has been much work on GP surrogate modeling with qualitative factors. This includes \citet{qian2008gaussian}, who proposed two types of kernels for correlation modeling with qualitative variables, both based on the Hamming distance. \citet{zhou2011simple} further introduced a simpler inference procedure via hypersphere decomposition. \citet{deng2017additive} explored additive GPs that permit different correlation structures between qualitative variables at varying combination levels. A recent work \citep{zhang2020latent} proposed a latent variable GP that maps qualitative factors onto a latent quantitative space, from which a standard GP for quantitative variables can be applied.

Despite this rich literature on GP modeling for qualitative factors, there has been considerably less work on the design of such experiments. A careful experimental design is essential in our path planning application, where each run is costly and many qualitative factors are present. For initial design, one reasonable choice is an orthogonal array \citep{hedayat2012orthogonal}, which provides run size economy with many factors and/or many factor levels. OA designs have been used for various GP models with qualitative factors; see, e.g., \cite{deng2015design,peng2019design}. One limitation, however, is that such designs can only be constructed for specific run sizes $n$, which may be restrictive given the expensive nature of computer experiments. To address this, recent promising work \citep{stokes2023metaheuristic} explores the use of metaheuristics for design optimization. The use of optimization heuristics, however, introduces two potential drawbacks. First, such heuristics do not exploit structure in the design formulation, and can thus be time-consuming for large problems with many factors and/or factor levels (e.g., path planning); we shall see this later. Second, using heuristics, it is difficult to gauge the optimization gap between the current solution and the desired global optimum. This in turn makes it difficult to reliably stop such algorithms for design optimization.

We thus propose a new design optimization framework, called QuIP, which tackles the problem of experimental design for many Qualitative factors via Integer Programming (IP). QuIP builds upon the qualitative GP model in \cite{qian2008gaussian} with an exchangeable correlation function. For initial design, we leverage a maximin design criterion derived asymptotically from its D-optimal design. We then show that this maximin design problem can be carefully reformulated as a variant of the assignment problem \citep{kuhn1955hungarian}, which has been extensively studied in operations research. With promising recent developments, state-of-the-art IP solvers can solve such problems to optimality \citep{mehbali2024solution}, and provide guidance on its optimization gap as the solver progresses. Our implementation of QuIP leverages the \texttt{Gurobi} solver \citep{gurobi} for efficient initial design optimization. For sequential design (specifically, for active learning or black-box optimization), we show that its design criterion under the above GP can similarly be formulated as a variant of the assignment problem. Similar solvers can then be used within QuIP for efficient and reliable sequential design optimization. We demonstrate the effectiveness of QuIP over existing design methods in a suite of path planning experiments and an application to rover trajectory optimization.



The paper is structured as follows. Section~\ref{sec:background} reviews the GP with qualitative factors in \cite{qian2008gaussian}, along with existing design methods and its potential limitations given many qualitative factors and/or levels. Section~\ref{sec:IP} presents the QuIP framework for initial design optimization and its reformulation as an assignment problem. Section~\ref{sec:sequential} extends this framework for sequential design. Section~\ref{sec:exp} compares QuIP to existing design methods in a suite of simulations and path planning experiments. Section~\ref{sec:DA} investigates its application for rover trajectory optimization. Section~\ref{sec:conc} provides concluding thoughts.

\section{Background \& Motivation}
\label{sec:background}
We first introduce the GP with qualitative factors in \cite{qian2008gaussian}, then review existing design approaches and its limitations given many qualitative factors and/or levels.

\subsection{GP modeling with qualitative variables}
\label{sec:krig}

Let $\mathbf{x} \in \mathcal{X}$ denote the considered $d$ qualitative factors, where $\mathcal{X} = \times_{l=1}^d [M]$ is its combinatorial parameter space and $[M] := \{1, \cdots, M\}$ is the set of $M$ levels for each factor. In what follows, we presume that the qualitative factors in $\mathbf{x}$ are categorical, i.e., the levels in $[M]$ are not ordered. This aligns with our path planning application, where each path decision is typically categorical in nature. 
Let $f(\mathbf{x})$ denote the scalar output of the computer simulator at parameters $\mathbf{x}$. We then model $f$ via a Gaussian process \citep{gramacy2020surrogates}:
\begin{equation}
    f(\cdot)\sim \mathcal{GP}\left\{\mu,\tau^2 \gamma_{\bs{\theta}}(\cdot,\cdot)\right\}.
    \label{eq:gp}
\end{equation}
Here, $\mu$ and $\tau^2$ are the mean and variance parameters of the GP, and $\gamma_{\bs{\theta}}(\cdot,\cdot)$ its correlation function with parameters $\bs{\theta}$. The specification of $\gamma_{\bs{\theta}}$ for qualitative factors is discussed later.


Next, suppose data $\mathbf{f}_n = (f(\mathbf{x}_1), \cdots, f(\mathbf{x}_n))$ are observed on $f$ at design points $\mathcal{D}_n = \{\mathbf{x}_1, \cdots, \mathbf{x}_n\}$. We assume that such data are observed without noise (i.e., the simulator is deterministic); the following framework extends naturally for Gaussian noise (see \citealp{gramacy2020surrogates}). Conditional on such data, the posterior distribution of $f$ at a new point $\mathbf{x}_{\rm new}$ follows:
\begin{equation}
    f(\mathbf{x}_{\rm new}) \mid \mathbf{f}_n \sim \mathcal{N}\left\{ \mu_n(\mathbf{x}_{\rm new}) ,  \sigma^2_n(\mathbf{x}_{\rm new})\right\},
\label{eq:gppred1}
\end{equation}
where its conditional mean and variance are given by:
\begin{align}
\begin{split}
    \mu_n(\mathbf{x}_{\rm new}) &= \mu + \mathbf{\gamma}_{\bs{\theta}}(\mathbf{x}_{\rm new},\mathcal{D}_n) \mathbf{\Gamma}_{\bs{\theta}}^{-1}(\mathcal{D}_n)(\mathbf{f}_n-\mu\mathbf{1}_n)\\   \sigma^2_n(\mathbf{x}_{\rm new})&=\tau^2\{1 - \mathbf{\gamma}_{\bs{\theta}}(\mathbf{x}_{\rm new},\mathcal{D}_n) \mathbf{\Gamma}_{\bs{\theta}}^{-1}(\mathcal{D}_n) \mathbf{\gamma}_{\bs{\theta}}(\mathcal{D}_n, \mathbf{x}_{\rm new})\}.
    \end{split}
    \label{eq:gppred2}
\end{align}
Here, $\mathbf{\gamma}_{\bs{\theta}}(\mathbf{x}_{\rm new},\mathcal{D}_n) = [\gamma_{\bs{\theta}}(\mathbf{x}_{\rm new},\mathbf{x}_i)]_{i=1}^n$ is the correlation vector between design points and $\mathbf{x}_{\rm new}$, and $\mathbf{\Gamma}_{\bs{\theta}}(\mathcal{D}_n) = [\gamma_{\bs{\theta}}(\mathbf{x}_i,\mathbf{x}_j)]_{i,j=1}^n$ is the correlation matrix for design points. The closed-form predictive equations in \eqref{eq:gppred1} and \eqref{eq:gppred2} facilitate effective use of the GP surrogate for downstream decision-making \citep{miller2024targeted,chen2022adaptive,song2023ace}.

A careful specification of the kernel $\gamma_{\bs{\theta}}$ is needed for GP surrogates with qualitative factors. Several flavors of kernels have been proposed in the literature, each adopting different structures on the underlying correlation function. This includes the unrestrictive kernel in \cite{qian2008gaussian} of the form $\gamma_{\bs{\theta}}(\mathbf{x},\mathbf{x}') = \prod_{l=1}^d \tau^{[l]}_{x_l,x_l'}$ where $\mathbf{T}^{[l]} = [\tau^{[l]}_{m,m'}]_{m,m'=1}^{M}$ is a correlation matrix with unit diagonal entries. The many parameters in this unrestrictive kernel, however, can be difficult to estimate with limited sample sizes. \cite{zhou2011simple} explored a hypersphere parametrization that reduces estimation costs. \cite{qian2008gaussian} further suggested the so-called exchangeable kernel:
\begin{equation}
\gamma_{{\rm E},\bs{\theta}}(\mathbf{x},\mathbf{x}') = \exp\left\{-\sum_{l=1}^d \theta_l\mathbf{1}(x_{l}\neq x_{l}')\right\}, \quad \bs{\theta} = (\theta_1, \cdots, \theta_d), \quad \theta_l > 0.
\label{eq:exchange}
\end{equation}
Such a kernel decreases the correlation between $f(\mathbf{x})$ and $f(\mathbf{x}')$ for every differing entry between $\mathbf{x}$ and $\mathbf{x}'$. The exchangeable kernel is thus well-suited for categorical factors with unordered levels; a similar kernel was used in \cite{joseph2007functionally}. Finally, a recent paper \citep{zhang2020latent} investigates a two-stage approach, where the qualitative factors are first mapped to a latent continuous space, then a standard GP for continuous variables is fit on such a space.

While many kernel choices are available, we adopt in the following the exchangeable kernel \eqref{eq:exchange} in \cite{qian2008gaussian} for two reasons. First, for our motivating path planning application, there are typically many qualitative factors (i.e., large $d$) that parametrize the path trajectory. Coupled with a small number of expensive simulation runs (i.e., small $n$), it is preferable to have a small number of kernel parameters for $\gamma_{\bs{\theta}}$. One can show that the exchangeable kernel has $d$ such parameters, whereas the unrestrictive kernel and the latent mapping approach require $dM(M-1)/2$ and $2d(M-3)$ parameters, respectively; the first is thus more preferable since it uses less parameters. Second, the simplicity of the exchangeable kernel \eqref{eq:exchange} is key for formulating the initial and sequential design problems as known integer programs that can be efficiently solved; this is outlined in Sections \ref{sec:IP} and \ref{sec:sequential}.

\subsection{Existing design methods}
\label{sec:exist}



While much has been done on GPs with qualitative factors, there has been relatively little work investigating its experimental design. Consider first the initial design of $\mathcal{D}_n$. A reasonable choice employed in the literature is an orthogonal array (OA; \citealp{hedayat2012orthogonal}). Recall that an OA design with strength $\lambda$ ensures every combination of $\lambda$ factors appears an equal number of times in the design. OA designs have been used in various GP models with qualitative factors, including mixed variable modeling \citep{deng2015design} and order-of-addition experiments \citep{peng2019design,lin2019order}. OA-based Latin hypercube designs \citep{tang1993orthogonal} have also gained popularity for GPs with continuous variables. There is, however, a key limitation of such designs for our setting: they can only be constructed for specific run sizes $n$. This can be restrictive for computer experiments \citep{owen1992orthogonal}, where each run is expensive and the computational budget is limited. 


To illustrate this run size constraint, consider the greedy-snake path planning problem in Section \ref{sec:snake}, which we discuss in greater detail later. Here, the goal is to navigate an object through a map to maximize cumulative reward, where its path is parametrized by $d=12$ factors with $M=5$ decisions (i.e., levels) for each factor. In this set-up, the smallest possible OA requires $n=100$ runs. Given the expensive nature of each run, using this OA for initial design can thus demand a large computational commitment for initial exploration, particularly since many more runs are needed for response surface optimization. We show later that, using an initial design that permits flexible (and smaller) run sizes, one can achieve good path planning performance with a budget of $n=100$ runs for this problem.

To bypass this run size constraint, there has been promising recent work on leveraging metaheuristic algorithms for design construction. This includes the work of \cite{stokes2023metaheuristic}, which utilized evolutionary \citep{das2010differential} and particle swarm optimization \citep{kennedy1995particle} algorithms for optimizing designs with many qualitative factors. The key appeal of such algorithms is that it permits \textit{flexible} design optimization for general run sizes $n$. However, one drawback with metaheuristic design algorithms is that they may take a long time to converge to a good solution \citep{abdel2023developments}, as they do not exploit structure in the optimization problem. This cost is exacerbated when there are many factors (i.e., large $d$) or many levels within a factor (i.e., large $M$); we shall see this later. Another drawback is that it may be difficult to gauge the optimization gap of the current solution (i.e., how far it is from the global optimum), as such heuristics typically do not exploit optimization duality \citep{luenberger1984linear}. As such, it may be difficult to reliably stop such heuristics for design optimization.

Consider next the sequential design of a subsequent point $\mathbf{x}_{n+1}$, given collected data $\mathbf{f}_n$ at initial design points $\mathcal{D}_n$. In computer experiments, sequential designs typically tackle one of two objectives \citep{gramacy2020surrogates}: they target either the improved prediction or optimization of $f$ over the parameter space $\mathcal{X}$. The first is known as \textit{active learning} in machine learning \citep{cohn1996active,mak2017information}, whereas the second is known as \textit{black-box optimization} \citep{frazier2018tutorial,miller2024diverse}. For GPs, there has been much work on developing criteria for selecting subsequent points. This includes the Active Learning Cohn and Mackay approaches \citep{cohn1996active,mackay1992information} for active learning, and the Expected Improvement and Upper Confidence Bound approaches \citep{Jones1998,auer2000using} for black-box optimization. However, even for the case of \textit{quantitative} factors, the optimization of such criteria can be difficult (see, e.g., \citealp{gramacy2022triangulation}); this thus becomes more challenging in the current case of 
\textit{qualitative} factors, as we shall see later. A promising recent work \citep{zhang2020latent} maps qualitative factors onto a latent quantitative space that is estimated from data, on which standard design criteria can be applied. With many qualitative factors and/or many levels, however, there can be much uncertainty in this estimated latent space with limited data, and its use for sequential design optimization may thus lead to suboptimal design points, as we show in later experiments.

We present next the proposed QuIP framework, which addresses the aforementioned challenges. The key idea is to reformulate both initial and sequential design problems as variants of assignment problems, which can be efficiently and reliably solved to optimality via state-of-the-art IP solvers. Using such reformulations as assignment problems, we show that QuIP enjoys improved initial and sequential design performance over existing methods in the setting of many qualitative factors, as motivated by our path planning application.

\section{QuIP: Initial Design Optimization}\label{sec:IP}

We now present the proposed QuIP approach for optimizing initial designs for GPs with qualitative factors. We first derive the considered maximin design criterion from its asymptotic D-optimal design. We then reformulate this problem as an assignment problem, and outline an iterative algorithm for efficient initial design optimization.

\subsection{Maximin design criterion}
\label{sec:maximin}

As before, let $\mathcal{D}_n = \{\mathbf{x}_1, \cdots, \mathbf{x}_n\}$ be the 
set of $n$ initial design points on $\mathcal{X}$. We consider the following maximin design formulation: 
\begin{equation}\label{eq:maximin}
  \mathcal{D}_n^* := \underset{\mathcal{D}_n \in \mathcal{X}^n}{\text{argmax}} \min_{i,i'} d_{\rm H}(\mathbf{x}_i,\mathbf{x}_{i'}).
\end{equation}
Here, $d_{\rm H}(\mathbf{x},\mathbf{x}') = \sum_{l=1}^d \mathbf{1}(x_l \neq x_l')$ is the Hamming distance \citep{hamming1986coding}, which returns the number of dissimilar entries between $\mathbf{x}$ and $\mathbf{x}'$. The maximin design $\mathcal{D}_n^*$ is thus a design that maximizes the number of dissimilar entries between any two points.


The following proposition justifies the use of the maximin design \eqref{eq:maximin} for a qualitative-input GP using the isotropic exchangeable kernel $\gamma_{{\rm E},\bs{\theta}}$ in \eqref{eq:exchange} with $\bs{\theta} = \theta \mathbf{1}$.
\begin{prop}{\text{[Asymptotic D-optimality]}}\label{prop:asyD}
    Let $\mathcal{D}_{n,\boldsymbol{\theta},k}^*$ be a D-optimal design under the scaled kernel $\gamma_{{\rm E},\bs{\theta}}^k$, i.e., it maximizes $\textup{det}\{\boldsymbol{\Gamma}_{\bs{\theta},k}(\mathcal{D}_n)\}$, where $\boldsymbol{\Gamma}_{\bs{\theta},k}(\mathcal{D}_n)$ is the covariance matrix of an $n$-point design $\mathcal{D}_n$ under the kernel $\gamma_{{\rm E},\bs{\theta}}^k$. Then, for a given run size $n$ and choice of correlation parameter $\theta > 0$ with $\boldsymbol{\theta} = \theta \mathbf{1}$, the maximin design $\mathcal{D}_n^*$ satisfies:
    \begin{equation}
    \frac{1+\textup{det}\left\{\boldsymbol{\Gamma}_{\bs{\theta},k}(\mathcal{D}^*_{n,\boldsymbol{\theta},k})\right\}}{1+\textup{det}\left\{\boldsymbol{\Gamma}_{\bs{\theta},k}(\mathcal{D}^*_n)\right\}}\leq 1+o_k(1).
    \end{equation}
    Here, $\zeta_k = o_k(1)$ if and only if, for any $\epsilon > 0$, $|\zeta_k| \leq \epsilon$ for $k$ sufficiently large.
\end{prop}
The proof of this proposition is provided in Appendix A.1, and relies on Claim 4.2 in \cite{johnson1990minimax}. From \eqref{eq:exchange}, note that $k$ acts as a scaling factor on the length-scale parameters $\theta_l$; a larger $k$ induces less correlation between two qualitative inputs $\mathbf{x}$ and $\mathbf{x}'$. Thus, this proposition shows that the relative gap in D-optimality between the D-optimal design $\mathcal{D}^*_{n,\bs{\theta},k}$ and the maximin design $\mathcal{D}_n^*$ becomes negligible as the modeled correlation from the GP decreases over all variables. Such arguments are known as \textit{scaling asymptotics} in the literature; see, e.g. \cite{lim2002design}.

From this, a natural question is why the maximin design $\mathcal{D}_n^*$ from \eqref{eq:maximin} should be used over the D-optimal design $\mathcal{D}_{n,\bs{\theta},k}^*$. One reason lies in the preference for space-filling designs over D-optimal designs for GPs with continuous inputs. Note that the latter designs are inherently model-dependent: they require the complete specification of the GP in terms of its kernel parameters $\bs{\theta}$, which are typically unknown prior to data. A misspecification of such parameters can lead to considerably poorer performance in practice. Space-filling designs, e.g., minimax \citep{mak2018minimax} and maximin \citep{morris1995exploratory} designs, bypass such an issue via a \textit{model-free} design criterion that relies solely on distances between points \citep{gramacy2020surrogates}. A similar reasoning can be used to justify the use of the maximin design \eqref{eq:maximin} as initial designs of GPs with qualitative factors.


\subsection{Formulation as an assignment problem}
\label{sec:mbap}
We now investigate how the maximin problem \eqref{eq:maximin} can be formulated as an assignment problem. The assignment problem is fundamental in operations research, with a rich and storied history \citep{kuhn1955hungarian,burkard2002selected, pentico2007assignment}. In its basic form, the assignment problem aims to assign $M$ jobs to $d$ workers such that its total utility is maximized. The following quadratic semi-assignment problem \citep{loiola2007survey, silva2021quadratic} will be of use to us later:
\begin{align}
\begin{split}
    \arg\max_{\{I_{jk}\}_{jk}} & \sum_{j=1}^d \sum_{j'=1}^d \sum_{k=1}^M \sum_{k'=1}^M I_{jk} I_{j'k'} u_{j'k} v_{jk'}\\
    \textup{s.t.} \quad & \sum_{k=1}^M I_{jk} = 1, \quad \text{ for all } j = 1, \cdots, d,\\
    & I_{jk} \in \{0,1\}, \quad \text{ for all } j = 1, \cdots, d \text{ and } k = 1, \cdots, M.
\end{split}
\label{eq:ap}
\end{align}
Here, $I_{jk}$ is a binary decision variable indicating if job $k$ is assigned to worker $j$, and $u_{jk}$ and $v_{jk}$ are known constants that quantify its resulting utility. The sum-to-one constraint in \eqref{eq:ap} ensures that each worker is assigned to exactly one job. The assignment problem has attracted much attention \citep{oncan2007survey}, with efficient optimization algorithms proposed in the literature \citep{kuhn1955hungarian, mills2007dynamic, date2016gpu}.

There are two variants of the assignment problem that will be of use later. The first is the \textit{bottleneck} assignment problem \citep{gross1959bottleneck}, which generalizes \eqref{eq:ap} by replacing the summed utility $\sum_{j}\sum_{j'} \sum_{k} \sum_{k'} I_{jk} I_{j'k'} u_{j'k} v_{jk'}$ with the bottleneck objective $\min_{j,j',k,k'} I_{jk} I_{j'k'} u_{j'k} v_{jk'}$. This is appropriate when one is interested in the worst-case assignment utility. Recent work has investigated optimization algorithms for such an extension \citep{punnen2011quadratic} with appealing theoretical guarantees \citep{burkard2011polynomially, abdel2018comprehensive}. The second variant is the \textit{multi-dimensional} assignment problem \citep{pierskalla1968multidimensional}, which generalizes \eqref{eq:ap} to account for additional assignment conditions. For example, in addition to assigning workers to jobs, one may wish to ensure assignment conditions on operating machines as well. Such an extension requires additional indices on the decision variables in \eqref{eq:ap}, with appropriate modifications on its objective via bottleneck or summation over its new indices. There has again been work on efficiently solving multi-dimensional assignment problems via IPs \citep{walteros2014integer, taha2014integer}. In what follows, we call an assignment problem with both variants a multi-dimensional bottleneck assignment problem (MBAP; \citealp{spieksma2000multi}); state-of-the-art IP solvers can be leveraged to efficiently solve MBAPs, as we show later.

We now show how the maximin design \eqref{eq:maximin} can be reformulated as an MBAP. Take a single design point $\mathbf{x}_i = (x_{i1},\cdots,x_{id}) \in \mathcal{X}$, which represents its $d$ qualitative factors. Consider a \textit{one-hot encoding} of $\mathbf{x}_i$ into the binary matrix $\mathbf{I}(\mathbf{x}_i) \in \{0,1\}^{ d\times M}$, where its $l$-th row is a binary encoding of its $l$-th factor level $x_{il}$. For example, suppose we have $M=3$ levels with $d=4$ factors. The one-hot encoding of the point $\mathbf{x} = (1, 2, 3, 1)$ takes the form:
\begin{equation}
\mathbf{I}(\mathbf{x}) = \begin{pmatrix}
    1 & 0 & 0 \\
    0 & 1 & 0\\
    0 & 0 & 1 \\
    1 & 0 & 0
\end{pmatrix},
\label{eq:dummy}
\end{equation}
where $(1,0,0)^\top, (0,1,0)^\top,$ and $(0,0,1)^\top$ are the binary encodings of levels 1, 2 and 3, respectively. The full design $\mathcal{D}_n$ with points $\mathbf{x}_1, \cdots, \mathbf{x}_n$ can then be represented by a similar one-hot encoding into the binary tensor $\mathcal{I}(\mathcal{D}_n) \in \{0,1\}^{n \times d \times M}$, where its $i$-th slice on the first dimension equals $\mathbf{I}(\mathbf{x}_i)$.

With this in hand, the maximin design problem \eqref{eq:maximin} can be reformulated as follows:
\begin{prop}
Let $\mathcal{D}_n^*$ be the maximin design in \eqref{eq:maximin}. Then its one-hot encoding $\mathcal{I}(\mathcal{D}_n^*)$ is the optimal solution to the assignment problem with decision variables $\mathcal{I} = {{[I_{ijk}]_{i=1}^n}_{j=1}^d}_{k=1}^M$:
\begin{align}
\begin{split}
\arg\max_{\mathcal{I}} & \min_{i,i'} \left\{ d - \textup{tr} (\mathbf{I}_i \mathbf{I}_{i'}^\top) \right\}\\
\textup{s.t.} \quad & \sum_{k=1}^M I_{ijk} = 1, \quad \text{ for all $i = 1, \cdots, n$ and $j = 1, \cdots, d$},\\
&I_{ijk} \in\{0,1\}, \quad \text{ for all $i$, $j$ and $k$,}
\end{split}
\label{eq:quipip}
\end{align}
where $\mathbf{I}_i = {[I_{ijk}]_{j=1}^d}_{k=1}^M$ is the $i$-th slice of the tensor $\mathcal{I}$ over its first dimension.
\label{prop:initialip}
\end{prop}
The proof of Proposition \ref{prop:initialip} is provided in Appendix A.2. Here, the binary variable $I_{ijk} = 1$ if, for the $i$-th design point, its $j$-th factor is set at level $k$; otherwise, $I_{ijk} = 0$. The sum-to-one constraint in \eqref{eq:quipip} ensures each factor is assigned to exactly one level. In the objective of \eqref{eq:quipip}, note that $\text{tr}(\mathbf{I}_i\mathbf{I}_{i'}^T)$ is of a summed quadratic form, and its minimization over indices $i$ and $i'$ is of a bottleneck form; thus, the formulation \eqref{eq:quipip} can be viewed as a multi-dimensional bottleneck assignment problem.

With this desirable MBAP structure, one can then leverage state-of-the-art IP solvers (e.g., \texttt{Gurobi}, which we use later) for efficient design optimization. IP solvers have undergone dramatic improvements, and are now widely used for decision-making in large-scale applications, e.g., flight scheduling \citep{alidaee2009note} and game scheduling for professional sports \citep{kendall2010scheduling}. Here, \texttt{Gurobi} offers two key advantages for design optimization. First, it provides a sophisticated framework that automatically identifies useful structure to exploit for quick optimization \citep{achterberg2020presolve}. Second, as the solver progresses, it provides estimates on the global optimum value. This is facilitated by an analysis of its dual problem \citep{wolsey2020integer}, which can be nicely formulated for the assignment problem and its variants \citep{guignard1989improved}. Such estimates permit \textit{reliable} design optimization by ensuring that the optimized designs are sufficiently close to the desired global optimum. We illustrate such advantages later in numerical experiments.



\subsection{Optimization via iterative feasibility programs}\label{sec:opt}

In practice, we find that a slight modification of the initial design assignment problem \eqref{eq:quipip} permits quicker optimization. Such a modification involves decomposing \eqref{eq:quipip} into an iterative sequence of feasibility programs. Let $q > 0$ be an auxiliary variable for the objective $\min_{i,i'} \left\{ d - \text{tr} (\mathbf{I}_i \mathbf{I}_{i'}^\top) \right\}$. One can then solve the assignment problem \eqref{eq:quipip} by solving the following feasibility programs iteratively:
\begin{align}
\begin{split}
\arg\max_{\mathcal{I}} & \quad 1\\
\textup{s.t.} \quad &  d - \text{tr} (\mathbf{I}_i \mathbf{I}_{i'}^\top) \geq q, \quad \text{ for all $i,i' = 1, \cdots, n$,}\\
& \sum_{k=1}^M I_{ijk} = 1, \quad \text{ for all $i = 1, \cdots, n$ and $j = 1, \cdots, d$},\\
&I_{ijk} \in\{0,1\}, \quad \text{ for all $i$, $j$ and $k$.}
\end{split}
\label{eq:quipipit}
\tag{FP}
\end{align}
Such a problem is known as a feasibility program \citep{bertsimas1997introduction}: it gauges whether a feasible solution exists given its constraints, and returns a so-called ``infeasibility certificate'' \citep{farkas1902theorie} if a solution cannot exist. Our idea is to iteratively solve \eqref{eq:quipipit} for increasing choices of $q$, until one finds a $q^*$ for which no feasible solution exists. 

There are three reasons why such an iterative approach can considerably speed up optimization. First (i), feasibility programs are generally easier to solve \citep{chinneck2007feasibility}, as the solver simply has to find an infeasibility certificate that guarantees a feasible solution cannot exist given the provided constraints. Second (ii), we show below that theoretical bounds can be used to cut down the range of $q$ that needs to be tested in the feasibility program \eqref{eq:quipipit}. Finally (iii), iterative optimization programs can often be solved efficiently in a sequential fashion via warm starts \citep{friedman2007pathwise}, which leverage solutions from previous iterations to speed up the solving of subsequent optimization problems. 

Let us consider point (ii) in greater detail. The following proposition provides a bound $q_0$ for which any choice of $q \leq q_0$ is guaranteed to yield a feasible solution for \eqref{eq:quipipit}.
\begin{prop}
Consider the feasibility program \eqref{eq:quipipit}. Define the bound:
\begin{equation}
q_0(n,d,M) = \max\left\{{k \in \{1, \cdots, d\} } : \frac{M^d}{\sum_{l=0}^{k-1}  
 {d \choose l} D_l} \geq n \right\},
\label{eq:lowerbd}
\end{equation}
where $D_l = l ! \sum_{i=0}^l {(-1)^i}/{i !}$.
Then, for any choice of $q \leq q_0(n,d,M)$, the feasibility program \eqref{eq:quipipit} has at least one feasible solution.
\label{prop:lowerbd}
\end{prop} The proof of Proposition \ref{prop:lowerbd} is provided in Appendix A.3. This proof relies on Theorem 4 of \citet{frankl1977maximum}, which provides bounds of the maximum size of a code.


The key use of this proposition is in reducing the range of $q$ that needs to be tested for the feasibility program \eqref{eq:quipipit}. This can be achieved as follows. Since the goal is to find the largest $q$ for which \eqref{eq:quipipit} is feasible, we begin by setting $q = q_0(n,d,M)$ and solving \eqref{eq:quipipit}, for which Proposition \ref{prop:lowerbd} guarantees a feasible solution. Next, we iteratively increase $q$ by one and re-solve \eqref{eq:quipipit}. This is repeated until \eqref{eq:quipipit} returns an infeasibility certificate from the \texttt{Gurobi} solver, which guarantees no feasible solution can be found for that choice of $q$. With this, the optimal $q^*$ is thus the largest $q$ for which a solution is found, and the optimal maximin design \eqref{eq:maximin} is its corresponding feasible solution.

A useful observation is that when $n \leq M$, i.e., the run size does not exceed the number of levels, this bound becomes $q_0(n,d,M)=d$. Conversely, when $n > M$, this bound can be shown to satisfy $q_0(n,d,M) \leq d-1$. This is not too surprising: when the run size exceeds the number of levels, it follows from the pigeonhole principle that, for each factor, there will be at least two runs at the same level. Thus, a Hamming distance of $d$ cannot be achieved for all pairwise runs, and hence $q_0$ must be less than $d$. This observation will be used later in the numerical experiments in Section \ref{sec:expinit} as an upper bound.

Consider next point (iii), the use of warm starts, which is widely employed for accelerating statistical learning algorithms, e.g., for variable selection \citep{friedman2007pathwise,mak2019cmenet} and matrix factorization \citep{zheng2024erpca}. Recall the above iterative strategy of solving the feasibility program \eqref{eq:quipipit} for increasing choices of $q$ (starting from $q_0$), until an infeasibility certificate is obtained. Here, warm starts involve the use of a found feasible solution of \eqref{eq:quipipit} at a previous iteration to initialize its optimization at the current iteration. Suppose, for a given $\tilde{q}$, \texttt{Gurobi} returns a feasible solution $\tilde{\mathcal{I}}$ for \eqref{eq:quipipit} with $q = \tilde{q}$. With this, the subsequent optimization of \eqref{eq:quipipit} with $q = \tilde{q}+1$ can be initialized at $\tilde{\mathcal{I}}$. While $\tilde{\mathcal{I}}$ may not be feasible for the new optimization problem, it provides the solver with a ``warm'' initialization for finding potential feasible solutions. This custom initialization can easily be implemented within \texttt{Gurobi} solvers. In later implementation, such warm starts can considerably accelerate the iterative optimization of \eqref{eq:quipipit} for increasing $q$.

\subsection{Algorithm statement}
\label{sec:alg}

\begin{algorithm}[!t]
\caption{QuIP: Optimizing a Maximin Initial Design}\label{alg:maximin}
\begin{algorithmic}
\State \textbf{Inputs}: Sample size $n$, number of qualitative factors $d$, number of levels $M$.
\State $\bullet$ Compute the bound $q_0(n,d,M)$ from \eqref{eq:lowerbd}, and set $\tilde{q} \leftarrow q_0(n,d,M)$.
\State $\bullet$ Find a feasible solution $\tilde{\mathcal{I}}$ (guaranteed to exist by Proposition \ref{prop:lowerbd}), by solving the feasibility program \eqref{eq:quipipit} with $q = \tilde{q}$ in \texttt{Gurobi}.
\While{$\tilde{q}\leq d-1$}
    \State $*$ Solve \eqref{eq:quipipit} with $q = \tilde{q}$ via \texttt{Gurobi}, with a warm start initialization $\mathcal{I} = \tilde{\mathcal{I}}$.
    
    \State $*$ Set \texttt{if.flg = 1} if the solver returns an infeasibility certificate. If not, set \texttt{if.flg = 0}, and update $\tilde{\mathcal{I}}$ as its found solution.
    \If{\texttt{if.flg = 1}} \textbf{break}
    \Else \; Increment $\tilde{q} \leftarrow \tilde{q} +1$.
    \EndIf
\EndWhile
\State \textbf{Outputs}: Optimized design $\mathcal{I}^* = \tilde{\mathcal{I}}$, maximized objective $q^* = \tilde{q}-1$.
\end{algorithmic}
\end{algorithm}

Algorithm~\ref{alg:maximin} summarizes the full initial design optimization algorithm. Given required inputs $n$, $d$ and $M$, we first evaluate the bound $\tilde{q} = q_0(n,d,M)$ from \eqref{eq:lowerbd}. We then solve the feasibility program \eqref{eq:quipipit} with $q = \tilde{q}$ to obtain a feasible solution $\tilde{\mathcal{I}}$; such a solution is guaranteed to exist by Proposition \ref{prop:lowerbd}. For this, no custom initialization is provided; we rely on the default initialization from \texttt{Gurobi}. Next, we increment $\tilde{q}$ by one, and re-solve the feasibility program \eqref{eq:quipipit} with $q = \tilde{q}$, using the previous solution $\tilde{\mathcal{I}}$ as a warm start. If an infeasibility certificate is returned by \texttt{Gurobi}, then the algorithm terminates; the optimal maximin design is the previous solution $\tilde{\mathcal{I}}$. If a feasible solution is found, then $\tilde{q}$ is incremented by one, and the procedure repeated until an infeasibility certificate is returned.


\section{QuIP: Sequential Design Optimization}\label{sec:sequential}
Next, we extend the QuIP framework for optimizing sequential designs for GPs with qualitative factors. We show next that, for the GP with an exchangeable covariance kernel, its sequential design formulation for both active learning and black-box optimization takes the form of an assignment problem that can be efficiently optimized.


\subsection{Sequential design for active learning}
\label{sec:ALM}

Consider first the goal of active learning, which targets the \textit{prediction} of $f$ over the parameter space $\mathcal{X}$. Suppose initial data are collected at design points $\mathcal{D}_n = \{\mathbf{x}_1, \cdots, \mathbf{x}_n\}$, yielding data $\mathbf{f}_n = (f(\mathbf{x}_1), \cdots, f(\mathbf{x}_n))$. As before, let us adopt the GP model with the exchangeable kernel $\gamma_{{\rm E},\boldsymbol{\theta}}$ in \eqref{eq:exchange}. A popular active learning framework is the Active Learning Mackay approach (ALM; \citealp{mackay1992information}), which selects the next design point $\mathbf{x}_{n+1}$ as:
\begin{align}
\begin{split}
    \mathbf{x}_{n+1}^* := \arg\max_{\mathbf{x}} \sigma_n^2(\mathbf{x}) &= \arg\max_{\mathbf{x}} \tau^2 \left\{1 - \mathbf{\gamma}_{\bs{\theta}}(\mathbf{x},\mathcal{D}_n) \mathbf{\Gamma}_{\bs{\theta}}^{-1}(\mathcal{D}_n) \mathbf{\gamma}_{\bs{\theta}}(\mathcal{D}_n, \mathbf{x}) \right\}\\
    &=\arg\min_{\mathbf{x}} \mathbf{\gamma}_{\bs{\theta}}(\mathbf{x},\mathcal{D}_n) \mathbf{\Gamma}_{\bs{\theta}}^{-1}(\mathcal{D}_n) \mathbf{\gamma}_{\bs{\theta}}(\mathcal{D}_n, \mathbf{x}).
    \label{eq:alm}
    \end{split}
\end{align}
Here, the expression for the posterior variance $\sigma^2_n(\mathbf{x})$ follows from \eqref{eq:gppred2}, with notation defined therein. In words, ALM sequentially selects the design point $\mathbf{x}_{n+1}$ with greatest posterior variance from the fitted GP. Such a strategy thus aims to reduce predictive variance by sampling points that are most uncertain from the surrogate.

We employ the one-hot encoding representation from Section \ref{sec:IP} (see \eqref{eq:dummy}) to reformulate the ALM problem \eqref{eq:alm} as an integer program. This is given in the following proposition:
\begin{prop}
Consider the ALM solution $\mathbf{x}_{n+1}^*$ in \eqref{eq:alm}. Its one-hot encoding $\mathbf{I}(\mathbf{x}_{n+1}^*)$ is the optimal solution to the assignment problem with decision variables $\mathbf{I} = {[I_{jk}]_{j=1}^d}_{k=1}^M$:
\begin{align}
\begin{split}
\arg \min_{\mathbf{I}} & \sum_{\mathbf{k}\in[M]^d}\xi_{\mathbf{k}} \prod_{l=1}^d I_{lk_l}, \quad \mathbf{k} = (k_1, \cdots, k_d) \\
\textup{s.t.} \quad & \sum_{k=1}^M I_{jk} = 1, \quad \text{ for all $j = 1, \cdots, d$},\\
&I_{jk} \in\{0,1\}, \quad \text{ for all $j$ and $k$,}
\end{split}
\label{eq:quipseqalm}
\tag{ALM}
\end{align}
where the cost coefficients $\xi_{\mathbf{k}}$ are provided in Appendix A.4 for brevity.
\label{prop:seqalm}
\end{prop}
The proof of this proposition is provided in Appendix A.4.

While the reformulated integer program \eqref{eq:quipseqalm} may appear daunting, the key advantage is that it takes the form of the so-called $d$-adic assignment problem (see \citealp{silva2021quadratic}), for which there is a body of literature on efficient optimization algorithms. Furthermore, the cost coefficients $\xi_{\mathbf{k}}$ are highly sparse (i.e., there are only a few non-zero entries), which greatly accelerates the optimization of \eqref{eq:quipseqalm}. As before, instead of adopting specific optimization algorithms, we employ the state-of-the-art IP solver in \texttt{Gurobi} to optimize \eqref{eq:quipseqalm}. Such a solver can automatically identify desirable structure for efficient optimization, and can provide guidance on sequential design quality via estimates of its global optimum through its dual formulation. We will demonstrate such advantages in later experiments for guiding reliable sequential design.

\subsection{Sequential design for black-box optimization}\label{sec:UCB}

Consider next the goal of black-box optimization, where we wish to \textit{optimize} $f$ over the parameter space $\mathcal{X}$. Without loss of generality, we presume the maximization of $f$ below; the minimization of $f$ can easily be accommodated by maximizing $-f$.

As before, suppose data $\mathbf{f}_n$ are collected at initial design points $\mathcal{D}_n$. Adopt again a GP model with exchangeable kernel $\gamma_{{\rm E},\boldsymbol{\theta}}$ in \eqref{eq:exchange}. A popular sequential design criterion for maximizing a black-box function $f$ is the so-called upper-confidence-bound (UCB; \citealp{auer2000using}) criterion. UCB selects the next evaluation point $\mathbf{x}_{n+1}$ as:
\begin{equation}
    \mathbf{x}_{n+1}^* := \arg\max_{\mathbf{x}} \left\{ \mu_n(\mathbf{x})+\lambda \sigma_n(\mathbf{x}) \right\}.
    \label{eq:ucb}
\end{equation}
Here, $\mu_n(\mathbf{x})$ and $\sigma_n(\mathbf{x})$ are the GP posterior mean and standard deviation from \eqref{eq:gppred2}, and $\lambda > 0$ is an exploration parameter. Note that $\lambda$ controls the trade-off between exploration and exploitation \citep{gramacy2020surrogates}, which is fundamental in reinforcement learning \citep{sutton1999reinforcement}. A smaller choice of $\lambda$ encourages the selection of points with larger posterior mean (i.e., exploitation), whereas a larger choice encourages points with greater posterior variance (i.e., exploration). This exploration parameter is typically fixed in implementation; we use $\lambda = 2.96$ in later experiments, as suggested by the literature \citep{chen2024hierarchical}.

Similar to active learning, using the one-hot encoding from Section \ref{sec:IP}, we can reformulate the above UCB problem as an integer program. The following proposition outlines this:
\begin{prop}
Consider the UCB solution $\mathbf{x}_{n+1}^*$ in \eqref{eq:ucb}. Its one-hot encoding $\mathbf{I}(\mathbf{x}_{n+1}^*)$ is the optimal solution to the assignment problem with decision variables $\mathbf{I} = {[I_{jk}]_{j=1}^d}_{k=1}^M$:
\begin{align}
\begin{split}
\arg \max_{\mathbf{I}} &\; \sum_{\mathbf{k} \in [M]^{d} } \tilde{\xi}_{\mathbf{k}} \prod_{l=1}^d I_{l k_l}, \quad \mathbf{k} = (k_1, \cdots, k_d)  \\
\textup{s.t.} \quad & \sum_{k=1}^M I_{jk} = 1, \quad \text{ for all $j = 1, \cdots, d$},\\
&I_{jk} \in\{0,1\}, \quad \text{ for all $j$ and $k$,}
\end{split}
\label{eq:quipsequcb}
\tag{UCB}
\end{align}
where the cost coefficients $\tilde{\xi}_{\mathbf{k}}$ are provided in Appendix A.5 for brevity.
\label{prop:ucb}
\end{prop}
The proof of this proposition is provided in Appendix A.5. Again, the key advantage of the integer program reformulation \eqref{eq:quipsequcb} is that it takes the form of a $d$-adic assignment problem \citep{silva2021quadratic}, which permits efficient optimization algorithms. The cost coefficients $\tilde{\xi}_{\mathbf{k}}$ are similarly highly sparse, allowing for optimization speed-ups. This IP reformulation can thus enable efficient global optimization via state-of-the-art solvers in \texttt{Gurobi}; we show this in later experiments.


\subsection{Algorithm statement}

\begin{algorithm}[!t]
\caption{QuIP: Optimizing Sequential Designs}\label{alg:BO}
\begin{algorithmic}
\State \textbf{Inputs}: Initial design points $\mathcal{D}_n = \{\mathbf{x}_i\}_{i=1}^n$ and data $\mathbf{f}_n$, number of desired sequential samples $n_{\rm seq}$, black-box optimization flag \texttt{bo.flg}.

\For{$l = 1, \cdots, n_{\rm seq}$}
\State $\bullet$ Estimate GP parameters $\widehat{\boldsymbol{\Theta}}_l$ from data $(\mathcal{D}_{n+l-1},\mathbf{f}_{n+l-1})$ via maximum likelihood.
\If{\texttt{bo.flg} = 0}
\State $\bullet$ Using parameters $\widehat{\boldsymbol{\Theta}}_l$, solve \eqref{eq:quipseqalm} via \texttt{Gurobi} to obtain the next point $\mathbf{x}_{n+l}$.
\Else
\State $\bullet$ Using parameters $\widehat{\boldsymbol{\Theta}}_l$, solve \eqref{eq:quipsequcb} via \texttt{Gurobi} to obtain the next point $\mathbf{x}_{n+l}$.
\EndIf
\State $\bullet$ Sample the black-box function $f$ at point $\mathbf{x}_{n+l}$.
\State $\bullet$ Update $\mathcal{D}_{n+l} \leftarrow \mathcal{D}_{n+l-1} \cup \mathbf{x}_{n+l}$, $\mathbf{f}_{n+l} = [\mathbf{f}_{n+l-1}, f(\mathbf{x}_{n+l})]$.
\EndFor
    
\State \textbf{Outputs}: Design points $\mathcal{D}_{n+n_{\rm seq}}$ with corresponding data $\mathbf{f}_{n+n_{\rm seq}}$.
\end{algorithmic}
\end{algorithm}

Algorithm \ref{alg:BO} summarizes the full sequential design optimization procedure. Beginning with initial design points $\mathcal{D}_n = \{\mathbf{x}_i\}_{i=1}^n$ and corresponding data $\mathbf{f}_n$, we specify $n_{\rm seq}$, the number of sequential points desired, and \texttt{bo.flg}, a flag indicating whether black-box optimization is desired (with active learning otherwise). For the first sequential sample, we first estimate via maximum likelihood the GP model parameters (i.e., its mean $\mu$, variance $\tau^2$, and length-scales $\{\theta_l\}_{l=1}^d$) using the data on hand; details on this can be found in \cite{gramacy2020surrogates}. Then, with fitted parameters plugged into \eqref{eq:quipseqalm} or \eqref{eq:quipsequcb}, we use \texttt{Gurobi} to solve its corresponding integer program to obtain the next evaluation point $\mathbf{x}_{n+1}$. This procedure is then repeated until the desired number of sequential samples have been collected.

\section{Numerical Experiments}
\label{sec:exp}

We now investigate the performance of QuIP initial and sequential designs in a suite of numerical experiments. We first inspect the computational performance of initial designs in simulations, then explore the effectiveness of sequential designs in two path planning applications: the ``maze-solving'' and ``greedy-snake'' problems. All computations are performed on a Intel Xeon Gold 6252 cluster, with 4 CPU cores and 4 GB of memory.

\subsection{Initial design}
\label{sec:expinit}
We inspect in the following the performance of QuIP initial designs. Here, a main benchmark for QuIP is the metaheuristic design optimization approach in \cite{stokes2023metaheuristic}, applied on the maximin design formulation \eqref{eq:maximin}. In our implementation, we employ the differential evolutionary (DE) approach in \cite{stokes2023metaheuristic} (using the author's code, with appropriate modifications to optimize \eqref{eq:maximin}), which performed the best of the compared metaheuristics in their paper. As such an approach is stochastic, we replicate this procedure 100 times to capture simulation variability. We include uniformly sampled random designs as another benchmark. QuIP initial designs are optimized using Algorithm \ref{alg:maximin}. All approaches are then compared on its design quality (in terms of the maximin criterion in \eqref{eq:maximin}) as well as its computation time. This is tested for a run size of $n=50$, with the number of factor levels $M$ varying from 10 to 30 and the number of factors $d$ varying from $M+5$ to $M+25$. Such large choices of $M$ and $d$ reflect our path planning application, where the number of decision variables $d$ and number of actions $M$ can be large.



\begin{figure}[!t]
    \centering
    \includegraphics[width=0.9\textwidth]{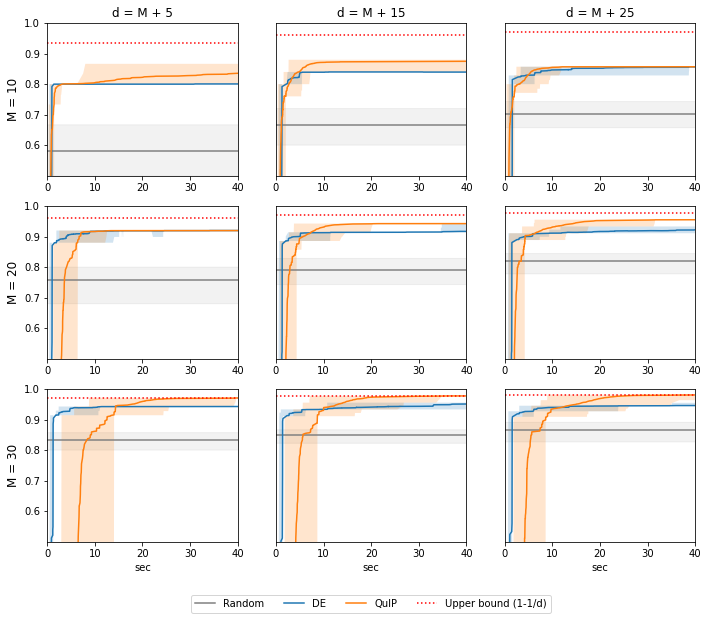}

    \caption{Plots of the maximin criterion in \eqref{eq:maximin} (scaled by $d$) vs. computation time for different initial design optimization methods, with shaded regions showing its 95\% intervals. Here, the run size is fixed at $n=50$, with a varying number of factors levels $M$ and number of factors $d$. The horizontal dotted red line shows the upper bound discussed in Section \ref{sec:opt} (scaled by $d$).}

    \label{fig:N50}
\end{figure}

Figure~\ref{fig:N50} shows the maximin criterion from \eqref{eq:maximin} (scaled by $d$ so that it is between $[0,1]$), as a function of computation time for design optimization. There are several observations of interest. First, in most cases, we see that QuIP considerably outperforms DE (i.e., its designs yield a larger maximin criterion) after a short warm-up period of at most 15 seconds. This warm-up period is expected for IP solvers: they often require some pre-processing of the provided formulation to identify desirable structure prior to optimization. Second, the improvement gap between QuIP and DE generally grows wider as the number of factors $d$ or the number of levels $M$ increases. This confirms our earlier observation that metaheuristic algorithms may be time-consuming for high-dimensional design optimization, i.e., for large $d$ and/or $M$. The proposed QuIP, via its reformulation as an assignment problem, appears to exploit such structure well for effective design optimization in high dimensions. Finally, recall from earlier discussions on Proposition \ref{prop:lowerbd} that, when $n > M$, the largest objective from the maximin problem \eqref{eq:maximin} cannot exceed $d-1$. This upper bound of $d-1$ (after scaling by $d$) is shown as a dotted red line in Figure \ref{fig:N50}. We see that, for all cases with $M=30$ levels, QuIP achieves this upper bound, which suggests that its designs are indeed globally optimal. Note that such an upper bound is likely not achievable for \textit{all} cases with $n > M$. For example, for $M=10$ and $M=20$, where QuIP falls short of this bound, it may very well be the case that the optimal maximin design \eqref{eq:maximin} cannot reach this upper bound, rather than slow optimization convergence from QuIP.

\subsection{Sequential design}
Next, we explore the sequential design performance of QuIP for active learning and black-box optimization in two path planning applications: the ``maze-solving'' and ``greedy-snake'' problems. Both problems are sufficiently high-dimensional in the number of factors $d$ and/or the number of levels $M$, such that an exhaustive optimization of \eqref{eq:alm} for ALM or \eqref{eq:ucb} for UCB is infeasible. For active learning, we consider the following benchmarks:
\begin{itemize}
\item \textit{Random}: This benchmark uses QuIP initial designs (which were shown to be preferable earlier) with random uniformly-sampled points for sequential design.
\item \textit{Candidate}: This benchmark uses QuIP initial designs, with a Monte Carlo ``candidate set'' approach for sequential design. Here, $C$ random candidate solutions are sampled on $\mathcal{X}$, and the best-performing solution (in terms of the ALM objective in \eqref{eq:alm}) is then taken as the next point. For fair comparison, $C$ is specified such that the computation time for this candidate set approach is comparable to that for the QuIP sequential design. Details on this are provided within each experiment.
\item \textit{QuIP}: This is the proposed approach, with QuIP initial designs (Section \ref{sec:IP}) and QuIP sequential designs (Section \ref{sec:sequential}). All IP optimizations are performed via the \texttt{Gurobi} software \citep{gurobi}.
\end{itemize}

Similar benchmarks are used for black-box optimization, with the obvious modification of optimizing the UCB formulation \eqref{eq:ucb}. Here, we include the LVGP approach in \cite{zhao2022optimal} (with QuIP initial design) as a further benchmark for black-box optimization, as implemented via the authors' code. Compared with other benchmarks, the LVGP employs a more sophisticated GP surrogate that leverages latent embeddings of qualitative factors. The hope is that, even with the simpler GP with exchangeable kernel \eqref{eq:exchange}, QuIP may offer improved performance via an effective and efficient sequential design optimization with IPs.



\subsubsection{Maze-solving problem}

\begin{figure}[!t]
\centering
        \includegraphics[width=0.26\textwidth]{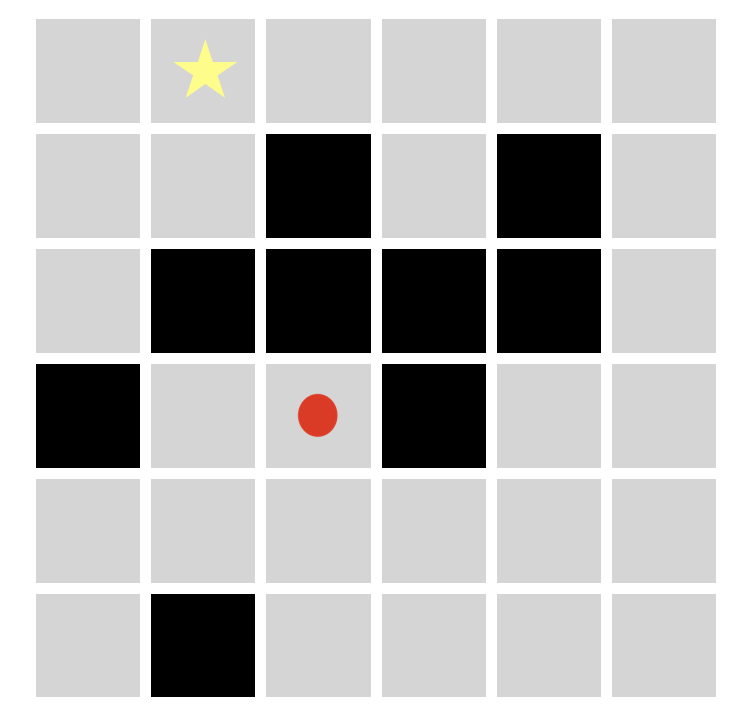}\quad\quad
           \includegraphics[width=0.26\textwidth]{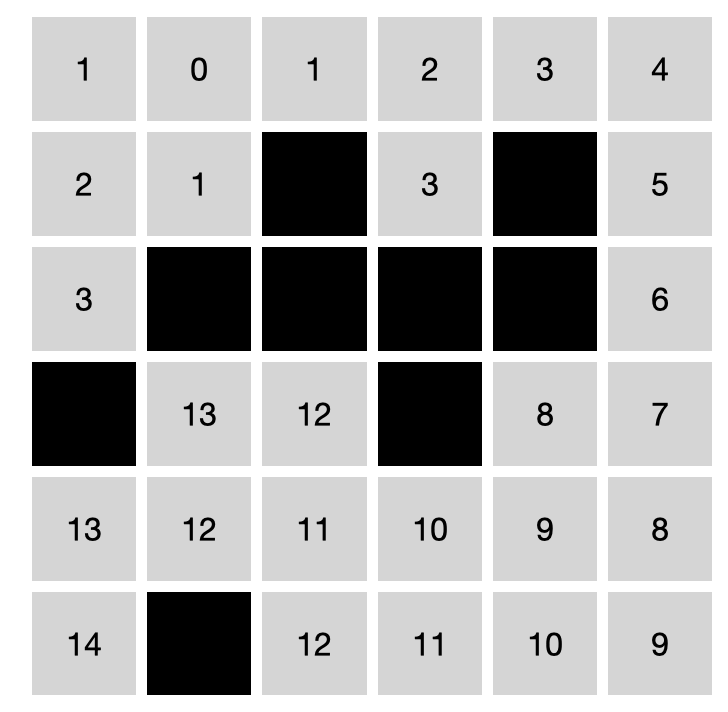}\quad\quad
        \includegraphics[width=0.26\textwidth]{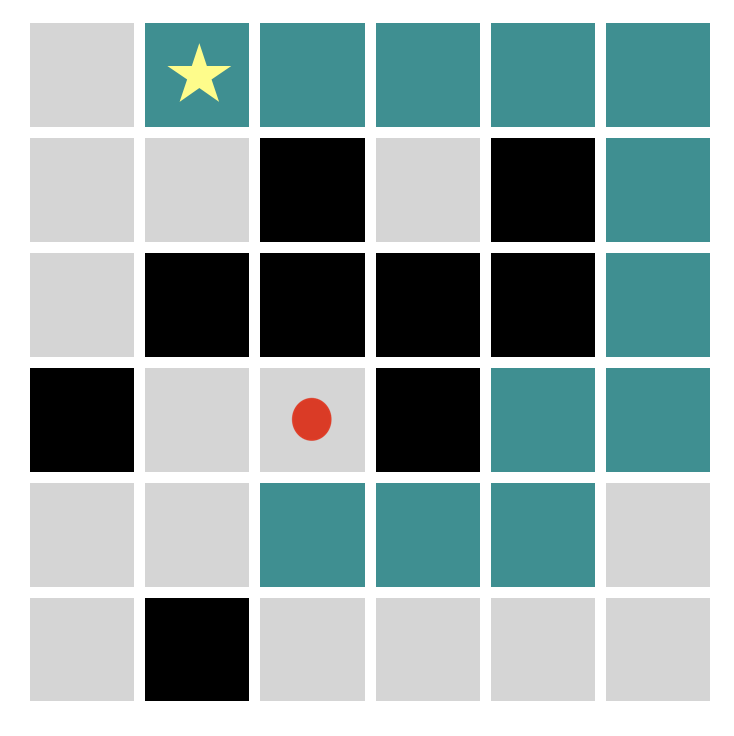}

    \caption{Visualizing the  maze-solving problem. [Left] The considered $6\times 6$ maze. The red dot and yellow star mark the start and desired end point of a path, respectively, with black squares marking obstacles that cannot be passed. [Middle] Each cell shows the cost of a path that ends on such a cell. [Right] An optimal path with minimal cost.}
    
    \label{fig:maze_illu}
\end{figure}

For the maze-solving problem, the goal is to navigate an object from start to end over a maze. In what follows, we adopt the 
$6 \times 6$ maze in Figure \ref{fig:maze_illu} (left), where certain grid cells serve as obstacles obstructing the path. At each step, the object has the option of moving one cell up, down, left, right or staying put; this amounts to $M=5$ choices at each step. We further consider paths of length $d=12$; this yields a massive search space of $5^{12} \approx 244$ million paths, which prohibits exhaustive search. The cost for a selected path is taken as its distance from the end of the path to the desired end point (see Figure \ref{fig:maze_illu} middle).

Here, the goals are to explore the cost response function over different paths, and to find an optimal path that minimizes such a cost; Figure \ref{fig:maze_illu} (right) shows one such path with minimal cost. These two goals can thus be explored sequentially via active learning and black-box optimization, respectively. For the latter goal, note that we are targeting the \textit{minimization} of $f$ here; this can easily be accommodated by maximizing its negative cost $-f$. All methods start with the QuIP initial design with $n=40$ points, and the simulation procedure is replicated 100 times to capture experiment variability.

\begin{figure}[!t]
    \centering
    \includegraphics[width=0.45\textwidth]{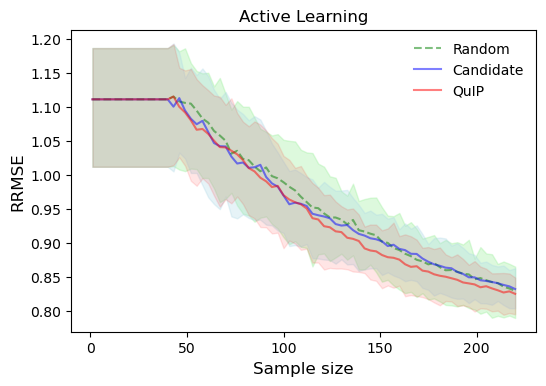}
    \includegraphics[width=0.45\textwidth]{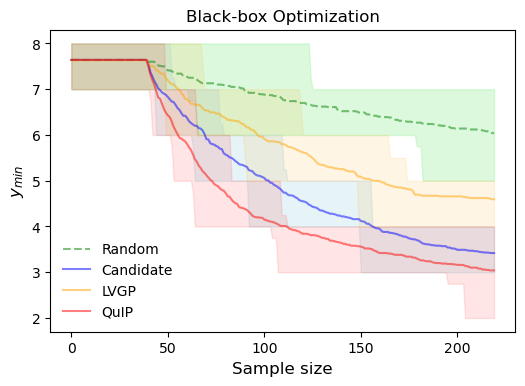}
    \caption{[Left] A plot of RRMSE vs. sample size, comparing various sequential design methods for active learning of the maze-solving problem. Here, shaded regions show its 95\% intervals. [Right] A plot of the current-best path cost (smaller-the-better) vs. sample size, comparing various sequential design methods for black-box optimization of the maze-solving problem.
    }
    \label{fig:maze_UCB}
\end{figure}

Consider first the use of sequential design for prediction. Figure~\ref{fig:maze_UCB} (left) compares the predictive performance of each method via its relative root-mean-squared-error (RRMSE) on a separate test set of $5,000$ random paths. Here, the RRMSE is computed as:
\begin{equation}
\text{RRMSE}=\sqrt{\frac{n_{\rm test}^{-1}\sum_{i=1}^{n_{\rm test}}(y_i-\widehat{y}_i)^2}{n_{\rm test}^{-1}\sum_{i=1}^{n_{\rm test}} (y_i-\bar{y})^2}},
\label{eq:rrmse}
\end{equation}
where $n_{\rm test}$ is the number of test points, $\bar{y}$ is the mean of the test responses, and $y_i$ and $\widehat{y}_i$ are the true and predicted responses for the $i$-th test point. For fair comparison, the candidate set size $C$ is selected such that it requires comparable computational cost as QuIP optimization; here, $C$ varies from $2.0 \times 10^4$ to $9.5 \times 10^4$, as sample size increases. There are several observations to note. First, compared to random sampling, QuIP offers considerable improvements with less variability, which shows the value of carefully-optimized points for active learning. Next, compared to the candidate set approach, QuIP similarly provides improved predictive performance. This shows the importance of solving the sequential design problem as a carefully-reformulated assignment program, compared to a Monte Carlo candidate set approach that may be inefficient for high-dimensional problems.


Consider next the use of sequential design for black-box optimization. Figure~\ref{fig:maze_UCB} (right) compares the optimization performance of each method via the cost of its current-best sampled path. Again, compared to the random and candidate approach (with $C$ ranging from $2.0 \times 10^4$ to $2.3\times 10^5$), QuIP offers considerably improved optimization performance with limited runs, thus showing the importance of reformulating the sequential design problem as an assignment problem for efficient optimization. QuIP similarly yields improvements over the LVGP approach. Recall that the latter makes use of a more sophisticated GP surrogate with latent embeddings of qualitative factors. This suggests that, even with the simpler GP using the exchangeable kernel \eqref{eq:exchange}, one can achieve improved sequential design performance with a structured optimization of the UCB acquisition via a careful reformulation as an assignment problem, thus demonstrating the importance of QuIP.


\subsubsection{Greedy-snake problem}
\label{sec:snake}

For the ``greedy snake'' problem, the goal is to navigate an object through a map, to maximize its cumulative reward by collecting prizes on each traversed square. In the following, we adopt the $8\times 8$ grid in Figure~\ref{fig:snakes_illu}, with prizes marked by yellow stars; its specific reward values are provided in Appendix B.1 for brevity. As before, the object can move up, down, left, right or stay put, amounting to $M=5$ choices at each step. Paths are constrained to have length $d=12$, which again yields a massive search space that prohibits exhaustive search. We further impose (i) a ``discounting'' reward mechanism over time to encourage collecting prizes at the start, (ii) a bonus mechanism for collecting prizes consecutively, (iii) an increasing penalty for not collecting prizes, and (iv) a penalty mechanism for going out of bounds. Details on this are provided in Appendix B.1. Figure~\ref{fig:snakes_illu} visualizes these rules. All methods start with the QuIP initial design with $n=40$ points, and the procedure is replicated 100 times to capture experiment variability.

\begin{figure}[!t]
\centering
        \includegraphics[width=0.3\textwidth]{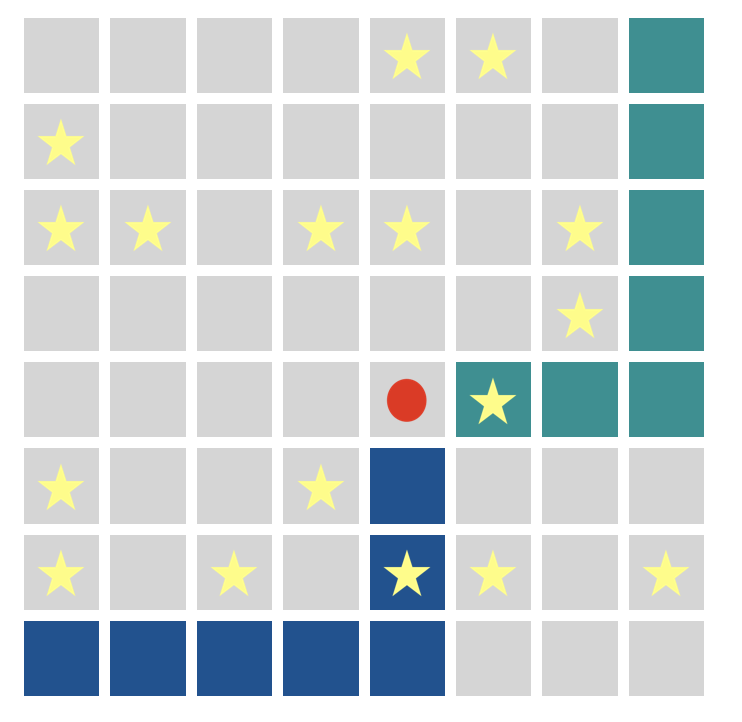}\quad
         \includegraphics[width=0.3\textwidth]{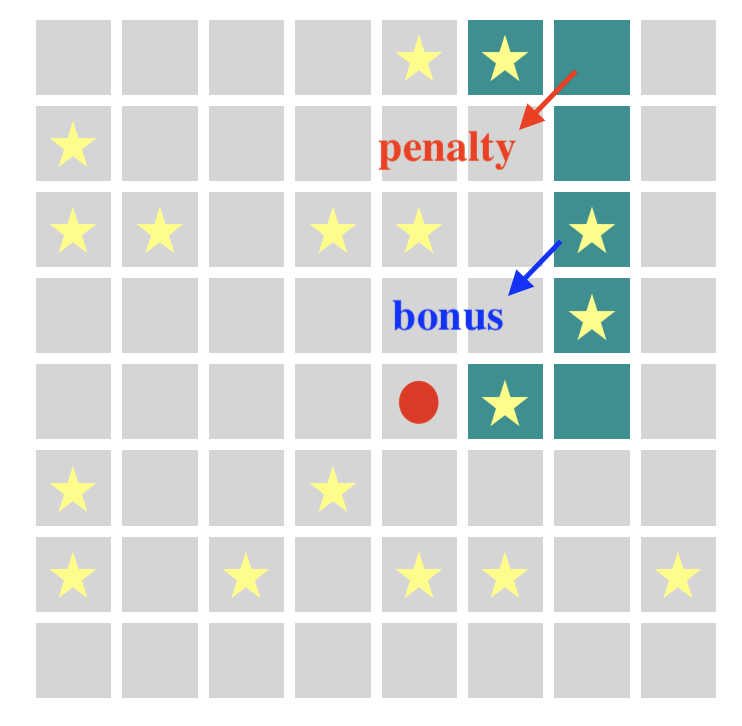}\quad
        \includegraphics[width=0.328\textwidth]{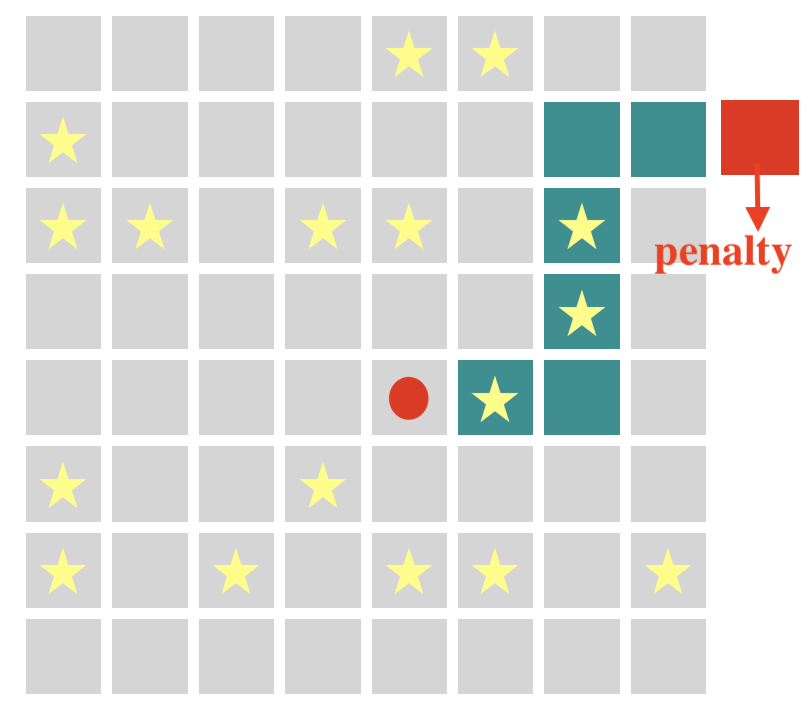}
  
    \caption{Visualizing the $8\times 8$ greedy-snake problem, where the red dot and yellow stars mark the starting point and prizes, respectively.  [Left] Visualizing the discounting reward mechanism: the reward from the green path is greater than that from the blue path as it is received earlier. [Middle] Visualizing the bonus and penalty mechanisms: the path receives a bonus at step 4 due to consecutive prizes, and a penalty at step 6 due to consecutive non-rewards. [Right] Visualizing the out-of-bounds penalty mechanism.}
    

    \label{fig:snakes_illu}
\end{figure}





\begin{figure}[!t]
    \centering

    \includegraphics[width=0.53\textwidth]{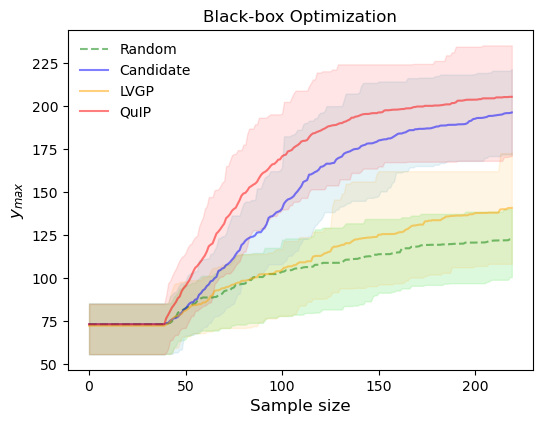}
            \includegraphics[width=0.4\textwidth]{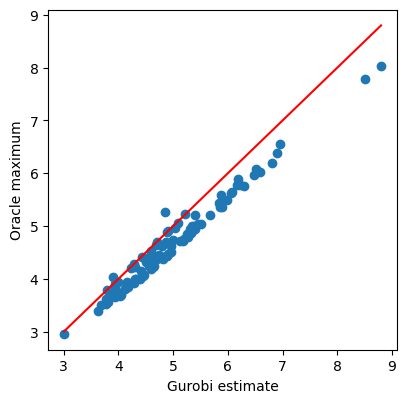}
    \caption{[Left] A plot of the current-best path reward (larger-the-better) vs. sample size, comparing various sequential design methods for black-box optimization of the greedy-snake problem. Here, shaded regions show its 95\% intervals. [Right] A scatterplot of the \texttt{Gurobi}-estimated global optima of \eqref{eq:quipsequcb} vs. its true oracle global optimum, for the 10-th sequential point in the greedy-snake problem. Here, each dot is for a single simulation replication, and the red line marks the ideal case where such estimates equal the oracle.
    }
    \label{fig:snakes_UCB}
\end{figure}

Figure \ref{fig:snakes_UCB} (left) compares the black-box optimization performance of each method via the cumulative reward of its current-best sampled path; here, the candidate approach is run with $C$ ranging from $1.5\times 10^4$ to $3.5\times 10^5$ (to match computation times for QuIP optimization). Compared to existing methods, QuIP finds paths that yield markedly greater cumulative rewards with limited runs. This again highlights the importance of reformulating the sequential design as an assignment problem for effective selection of subsequent runs. LVGP performs similarly to random designs at the beginning, but slowly improves as $n$ increases. This is not surprising, as the LVGP may need larger sample sizes to adequately estimate the required mapping to the latent quantitative space. To contrast, with a simpler GP model, QuIP again achieves improved black-box optimization performance via a structured optimization of the UCB acquisition as an assignment problem.

We further examine how \texttt{Gurobi} can guide reliable sequential design optimization via its inferred optimization gap. Recall from Section \ref{sec:mbap} that the global optimum for an assignment problem can be estimated via its dual formulation from \texttt{Gurobi}; this can be used to stop the QuIP design optimization once its current objective is sufficiently close to such an estimate. To assess the accuracy of such estimates, Figure~\ref{fig:snakes_UCB} (right) plots, for the 10-th sequential point in the greedy-snake set-up, the \texttt{Gurobi}-estimated global optimum for the UCB assignment problem \eqref{eq:quipsequcb} vs. its true oracle global optimum. For all simulation replications, we see that the \texttt{Gurobi} estimates are all close to the global optima, as expected. Moreover, nearly all estimates are conservative in that they slightly over-estimate the oracle optima; this is desirable, as it ensures the optimization does not stop pre-maturely before achieving the stopping condition. For QuIP optimization in our experiments, the solver is run until the difference between its current objective and the \texttt{Gurobi}-estimated optimum is within 10\% of this optimum, which ensures the QuIP-optimized design points are sufficiently high quality.



\section{Application: Rover Trajectory Optimization}\label{sec:DA}

Finally, we explore the performance of QuIP in a path planning application for rover trajectory optimization. This has broad applications in, e.g., robotics \citep{path_planning_robotics}, assembly planning \citep{path_assembly} and aircraft navigation \citep{path_planning_flight}. For sophisticated problems, however, the evaluation of a selected path can require costly computer simulations \citep{frank2008efficient}. For example, in navigating an exploration rover on Mars \citep{carsten_rover}, a detailed simulation of a planned rover path can require minutes of computing time to accurately capture its operating environment. 

\begin{figure}
    \centering
    
        \includegraphics[width=0.4\textwidth]{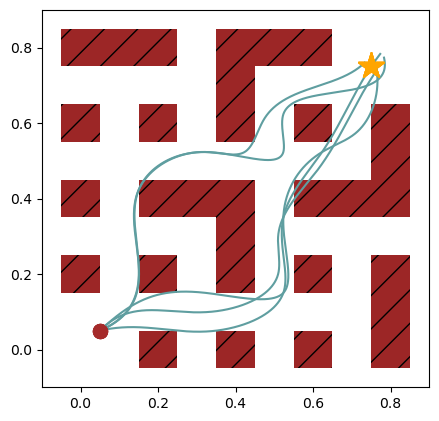}
          \includegraphics[width=0.55\textwidth]{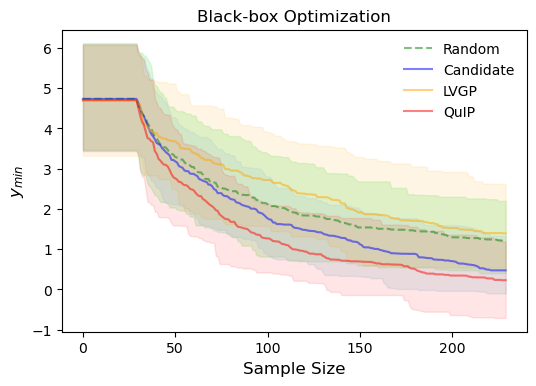} 
    \caption{[Left] Visualizing low-cost paths in the rover trajectory optimization application. Here, the red dot and yellow star mark the starting and desired end point, respectively. Brown blocks represent obstacles that can be passed through with greatly reduced speed. [Right] A plot of the current-best path cost (smaller-the-better) vs. sample size, comparing various sequential design methods for black-box optimization in the rover trajectory application.}
    \label{fig:rover}
\end{figure}

To investigate this problem, we adopt the rover optimization simulator in \cite{wang2018batched}, which simulates a rover passing through the obstacle course in Figure \ref{fig:rover} (left). Here, the brown blocks represent obstacles; these can be passed through but with greatly reduced speed. The goal is to navigate the rover from the start point (indicated by a red dot) to the target end point (yellow star) as efficiently as possible. The cost of a selected path is thus jointly determined from (i) the distance of its end point to the target end point, and (ii) the total time spent within an obstacle. Details on this cost function are provided in Appendix B.2. Following existing work on rover optimization \citep{miller2024diverse}, we consider a discretization of the path into $d=8$ equal timesteps. Within each timestep, the rover moves at a constant nominal velocity (before potential speed reductions from an obstacle) towards a pre-specified direction angle. We adopt the following discrete space of $M=9$ decisions at each timestep: the first eight corresponds to different combinations of the two speed choices (``low'' and ``high'') and four angle choices (0, $\pi/6$, $\pi/3$ and $\pi/2$ radians), and the last is to stay put with zero speed.

Simulations are performed via the Python path-planning module from \cite{wang2018batched}. Figure \ref{fig:rover} (left) shows examples of paths with low costs, which largely avoid any obstacles and end close to the target end point. While each simulation run here requires only seconds, the exploration of all $9^8 \approx $ 43 million possible paths via simulation can be prohibitively expensive for cost minimization. This thus provides a nice case study for exploring the effectiveness of QuIP. As before, we employ random designs, the candidate set approach (with $C$ ranging from $5.0\times 10^3$ to $1.5\times 10^5$ to match QuIP optimization times) and the LVGP \citep{zhang2020latent} as baseline methods. All methods begin with the same $n=30$ QuIP initial design, then progress with 200 sequential design points.



Figure~\ref{fig:rover} (right) shows the black-box optimization performance of each method, in terms of the cost of its current-best path over sample size. We observe similar observations as before. QuIP again outperforms all the competitors, thus showing the importance of optimizing sequential design points as an assignment problem. LVGP performs even worse than random sampling in this case; this may be due to its inability to learn the underlying latent structure with small run sizes in this high-dimensional problem.


\section{Conclusion}
\label{sec:conc}

In this work, we presented a new method, called QuIP, for designing experiments of GP surrogates with many qualitative factors via integer programming. This is guided by our motivating application in path planning, where there may be many discrete decision variables and the evaluation of a path can involve expensive simulations. QuIP builds upon the qualitative GP in \cite{qian2008gaussian} with an exchangeable correlation function. For initial design, we showed that a maximin design (derived asymptotically from its D-optimal design) can be reformulated as a variant of the assignment problem. For sequential design, we derived design criteria that can be similarly reformulated as assignment problems, for both active learning and black-box optimization. QuIP then leverages state-of-the-art integer programming algorithms to solve such formulations for efficient and reliable design optimization. Finally, we demonstrated the effectiveness of QuIP over existing design methods in a suite of path planning experiments and an application to rover trajectory optimization.

These promising results bring about several intriguing directions for future work. First, it would be interesting to investigate broader GP models on qualitative factors for which integer programming formulations can be similarly exploited for efficient experimental design. This includes recent work on order-of-addition experiments \citep{stokes2023metaheuristic} and hierarchically sparse GPs \citep{tang2024hierarchical}. Second, we are exploring extensions of integer programming sequential design methods for broader objectives, e.g., fault localization \citep{ji2024bayesflo} and multi-armed bandits \citep{slivkins2019introduction,mak2022tsec}. Finally, the application of such methods for discrete decision-making in public policy (e.g., \citealp{gilbert1995artificial}) has much potential; we are investigating this in ongoing collaborations.

\spacingset{1.0}
\setlength{\bibsep}{0pt plus 0.3ex}
\footnotesize
\bibliography{reference}

\newpage
\renewcommand{\appendixpagename}{Appendix}

\begin{appendices}
\section{Technical Proofs}\label{supp:1}

\subsection{Proof of Proposition 1}
\label{appendix:prop1}
Note that the isotropic exchangeable kernel $\gamma_{\mathbf{E},\bs{\theta}}(\mathbf{x},\mathbf{x}')$ with $\bs{\theta} = \theta \mathbf{1}$ can be written as:
\begin{equation}
    \gamma_{\mathbf{E},\bs{\theta}}(\mathbf{x},\mathbf{x}') = \exp\{-\theta \cdot d_{\rm H}(\mathbf{x},\mathbf{x}')\},
\end{equation}
where $d_{\rm H}(\cdot,\cdot)$ is the Hamming distance. This proposition is then proven by applying Claim 4.2 in \cite{johnson1990minimax}.

\subsection{Proof of Proposition 2}
\label{appendix:prop2}

Let $\mathcal{D}_n^*$ be the maximin design defined in Equation (5) of the main paper, and let $\mathbf{I}^*:=\mathcal{I}(\mathcal{D}_n^*)$ be its one-hot encoding representation. Define $q^*=d(\mathcal{D}_n^*)$ as the maximum minimum pairwise distance in the design $\mathcal{D}_n^*$. Proving Proposition 2 is thus equivalent to showing (i) $\max_{\mathcal{I}}\min_{i,i'}\{d-\textup{tr}(\mathbf{I}_i\mathbf{I}_{i'}^\top)\}=q^*$ and (ii) $\min_{i,i'}\{d-\textup{tr}(\mathbf{I}_i^*\mathbf{I}_{i'}^{*\top})\}=q^*$ under constraints listed in (FP) of the main paper.

First, note that for any two design points $\mathbf{x}_i$ and $\mathbf{x}_{i'}$ in $\mathcal{D}_n$, $d_{\rm H}(\mathbf{x}_i,\mathbf{x}_{i'})=\sum_{l=1}^d \mathbf{1}(x_{il}\neq x_{i'l})=d-\text{tr}\{\mathbf{I}(\mathbf{x}_i) \mathbf{I}(\mathbf{x}_{i'})^\top\}\leq d$. To prove (i), we use the bijective nature of the one-hot encoding to obtain:
\[
q^*=\max_{\mathcal{D}_n}\min_{i,i'}d_H(\mathbf{x}_i,\mathbf{x}_{i'}) = \max_{\mathcal{I}}\min_{i,i'}\{d-\textup{tr}(\mathbf{I}_i\mathbf{I}_{i'}^\top)\}.
\]
 To prove (ii), note that:
\begin{align*}
q^* & = \max_{\mathcal{D}_n^*}\min_{i,i'} d_{\rm H}(\mathbf{x},\mathbf{x}_{i'})\\
&=\max_{\mathcal{D}_n^*}\min_{i,i'}\{d-\text{tr}\{\mathbf{I}(\mathbf{x}_i)\mathbf{I}(\mathbf{x}_{i'})^\top\}\}\\
&=\min_{i,i'}\{d-\text{tr}(\mathbf{I}_i^*{\mathbf{I}_{i'}^*}^\top)\},
\end{align*}
which completes the proof.

\subsection{Proof of Proposition 3}
\label{appendix:prop3}

Denote $N(M,d,q)$ as the number of feasible designs with $d$ factors and $M$ distinct categories whose minimum pairwise Hamming distance is $q$. Further denote $P(M,q)$ as the number of permutation arrays consisting of elements $\{1,2,\cdots,M\}$ with minimum pairwise distance $q$. 

We will first show that $P(M,q)\leq N(M,d,q)$ when $M\leq d$. When $d=q$, it is clear that $P(M,q)\leq N(M,M,q)$, since permutation arrays are a special case of qualitative factor designs restricted to sampling without replacement. Further, we have $N(M,d,q)\geq N(M,d-1,q)$, since adding a new factor does not decrease the counts of different elements between design points. Thus, it follows that $P(M,q)\leq N(M,M,q)\leq N(M,d,q)$ for $M\leq d$.

Next, we make use of a lower bound of $P(M,q)$ in Theorem 4 of \citet{frankl1977maximum}:
\begin{equation}
 \frac{M^d}{\sum_{l=0}^{q-1}  
 {d \choose l} D_l}\leq  P(M,q),
\end{equation}
where $D_l=l ! \sum_{i=0}^l {(-1)^i}/{i !}$. Therefore, for any $q\in\{1,\cdots,d\}$ satisfying:
\begin{equation}
n\leq \frac{M^d}{\sum_{l=0}^{q-1}  
 {d \choose l} D_l },
\end{equation}
there exists a feasible solution of run size $n$ for (FP) in the main paper, since $n\leq P(M,q)\leq N(M,d,q)$. Finally, because $N(M,d,q)$ is a decreasing function of $q$, there exists a feasible design of run size $n$ of size $n$ for any choice of $q$ satisfying:
\begin{equation}
q\leq \max\left\{q\in\{1,\cdots,d\}\left| \frac{M^d}{\sum_{l=0}^{q-1}  
 {d \choose l} D_l } \geq n\right.\right\},
\end{equation}
which completes the proof.

 \subsection{Proof of Proposition 4}\label{appendix:prop4}
 Let $\mathbf{I}_{r}$ denote the $r$-th slice of the tensor $\mathbf{I}(\mathcal{D}_n)$ along its first dimension. Further let $\mathbf{I}\in\{0,1\}^{d\times M}$ be the one-hot encoding of the new point $\mathbf{x}$. The optimal ALM objective can then be written as:
\begin{align}
\begin{split}
    & \min_{\mathbf{x}}\mathbf{\gamma}_{\bs{\theta}}(\mathbf{x},\mathcal{D}_n) \mathbf{\Gamma}_{\bs{\theta}}^{-1}(\mathcal{D}_n) \mathbf{\gamma}_{\bs{\theta}}(\mathcal{D}_n, \mathbf{x})\\
    & \quad = \min_{\mathbf{I}}\sum_{r=1}^n\sum_{s=1}^n  \exp\left[-\sum_{k=1}^d \left\{1-\left(\mathbf{I}\mathbf{I}_{r}^\top\right)_{kk}\right\}\theta_k\right] \left[\bs{\Gamma}_{\bs{\theta}}^{-1}(\mathcal{D}_n)\right]_{rs} \exp\left[-\sum_{k=1}^d \left\{1-\left(\mathbf{I}\mathbf{I}_{s}^\top\right)_{kk}\right\}\theta_k\right] \\
    & \quad =
    \min_{\mathbf{I}}\sum_{r=1}^n\sum_{s=1}^n\left[\bs{\Gamma}_{\bs{\theta}}^{-1}(\mathcal{D}_n)\right]_{rs} \prod_{k=1}^d\left[1+\left\{1-\mathbf{I}\mathbf{I}_{r}^\top\right\}_{kk}\left(e^{-\theta_k}-1\right)\right] \left[1+\left\{1-\mathbf{I}\mathbf{I}_{s}^\top\right\}_{kk}\left(e^{-\theta_k}-1\right)\right]. 
    \end{split}
\end{align}
The last expression thus reduces to the sum of polynomials with terms $\prod_{l=1}^d I_{l k_l}$, which proves the proposition.


  \subsection{Proof of Proposition 5}\label{appendix:prop5}
  We show first that the posterior GP mean and variance of $\mathbf{I}$ (the one-hot encoding of the new point $\mathbf{x}$) takes a desirable form. Note that its posterior GP mean can be written as:
  \begin{align}
  \begin{split}
{\mu}_n(\mathbf{I})&= \mu + \sum_{r=1}^n\sum_{s=1}^n\exp\left[-\sum_{k=1}^q \left\{1-\left(\mathbf{I}\mathbf{I}_{r}^\top\right)_{kk}\right\}\theta_k\right] \{\bs{\Gamma}_{\bs{\theta}}(\mathcal{D}_n)^{-1}\}_{rs}(\mathbf{f}_{n,s}-\mu)\\
&= {\mu} + \sum_{r=1}^n\sum_{s=1}^n \{\bs{\Gamma}_{\bs{\theta}}(\mathcal{D}_n)^{-1}\}_{rs} (\mathbf{f}_{n,s}-\mu)\left(\prod_{k=1}^d \left[1+ \left\{1-\left(\mathbf{I}\mathbf{I}_{r}^\top\right)_{kk}\right\}\left(e^{-\theta_k}-1\right)\right]\right)\\
&=\widehat{\mu}_n + \sum_{\mathbf{k}\in[M]^d}\xi^\mu_{\mathbf{k}}\prod_{l=1}^d {I}_{lk_l},
\end{split}
  \end{align}
  for cost coefficients $\xi^\mu_{\mathbf{k}}$, where the last step follows from the derivation in Appendix \ref{appendix:prop4}. Similarly, its posterior GP variance can be written as: 
  \begin{align*}
      {\sigma}^2_n(\mathbf{I})&=\tau^2\left(1-\sum_{r=1}^n\sum_{s=1}^n\exp\left[-\sum_{k=1}^d\left\{\mathbf{I}\left(\mathbf{I}_{n,r}+\mathbf{I}_{n,s}\right)^\top\right\}_{kk}\theta_k\right] \{\bs{\Gamma}_{\bs{\theta}}(\mathcal{D}_n)^{-1}\}_{rs}\right)\\
      &=\tau^2\left(1-\sum_{\mathbf{k}\in[M]^d}\xi_{\mathbf{k}} \prod_{l=1}^d {I}_{lk_l}  \right),
  \end{align*}
for cost coefficients $\xi_{\mathbf{k}}$, where the last step again follows from Appendix \ref{appendix:prop4}.
With this, we can show:
  \begin{align*}
      {\mu}_n(\mathbf{I}) + \lambda {\sigma}_n(\mathbf{I})
      &={\mu} + \sum_{\mathbf{k}\in[M]^d}\xi^\mu_{\mathbf{k}}\prod_{l=1}^d \mathbf{I}_{lk_l}+\lambda {\tau}\sqrt{1-\sum_{\mathbf{k}\in[M]^d}\xi_{\mathbf{k}} \prod_{l=1}^d {I}_{lk_l}}\\
      &={\mu} +\lambda{\tau} +\sum_{\mathbf{k}\in[M]^d}\left(\xi^\mu_{\mathbf{k}}-{\tau}\lambda\sqrt{1-\xi_{\mathbf{k}}}\right)\prod_{l=1}^d {I}_{lk_l}\\
      &= C +  \sum_{\mathbf{k}\in[M]^d}\left(\xi^\mu_{\mathbf{k}}-{\tau}\lambda\sqrt{1-\xi_{\mathbf{k}}}\right)\prod_{l=1}^d {I}_{lk_l},
  \end{align*}
  for some constant $C$, which proves the desired proposition.

\section{Numerical Experiment Details}\label{supp:2}


\subsection{Greedy-snake problem rewards}
\label{sec:appsnake}
We consider the following reward set-up for the greedy-snake problem in Section 5.2.2 of the main paper. Here, the goal navigate an object through a map to maximize its cumulative reward $\sum_{j=1}^d R_j$, where $R_j$ is the reward given at step $j$. Let $\Omega=\{(x,y):x=1, \cdots, 8, y = 1, \cdots, 8\}$ denote each square in the $8\times 8$ grid plane, and let $\mathcal{S} \subseteq \Omega$ denote the subset of squares with prizes from Figure 4 of the main paper. Letting $(x_j,y_j)$ be the square traversed on step $j$, we define the reward at the $j$-th step as:
\begin{equation*}
    R_j=\left\{
    \begin{array}{ll}
   5(d-j+1)      &  \text{if } (x_j,y_j)\in \mathcal{S} \text{ and } (x_{j-1},y_{j-1})\notin \mathcal{S},\\
   10(d-j+1)     & \text{if } (x_j,y_j)\in\mathcal{S} \text{ and } (x_{j-1},y_{j-1})\in\mathcal{S},\\
   -2(j-1) & \text{if } (x_j,y_j)\notin \mathcal{S},\\
   -10 & \text{if } (x_j,y_j)\notin \Omega.
    \end{array}
    \right.
\end{equation*}
\subsection{Rover trajectory cost function}

Let $\mathbf{x}(t)$ be the location of the rover at time $t$, and suppose its path times are discretized as $t_1 = 0,t_2, \cdots, T$, where $T$ is its ending time. The cost function of its trajectory is then given as:
\begin{equation}
\sum_{j=1}^{T-1} \frac{c\left(\mathbf{x}(t_j)\right)+c\left(\mathbf{x}(t_{j+1})\right)}{2}\left\|\mathbf{x}(t_{j+1})-\mathbf{x}(t_j)\right\| + 50\left\|\mathbf{x}(T)-(0.75,0.75)\right\|_2-5,
\end{equation}
where $c(\mathbf{x})=30\cdot\mathbf{1}\left\{\mathbf{x} \text{ is on an obstacle}\right\}$ + 0.05.
\end{appendices}

\end{document}